\documentclass{appolb}
\usepackage{graphicx}
\usepackage{subfigure}

\preprint{}

\begin{document}

\title{{\bf Hydro-inspired parameterizations of freeze-out in
relativistic heavy-ion collisions}\thanks{Lecture presented  by WF at the 
{\em XLIV Cracow School of Theoretical Physics}, Zakopane, Poland, 28 May -- 6 June 2004}
\thanks{Research supported in part by the Polish State Committee for
Scientific Research, grant number 2~P03B~05925}}
\author{Wojciech Florkowski $^{1,2}$ and Wojciech Broniowski $^2$ 
\address{
$^1$ Institute of Physics, \'Swi\c{e}tokrzyska Academy, \\
ul.~\'Swi\c{e}tokrzyska 15, PL-25406~Kielce,~Poland \\
$^2$ Institute of Nuclear Physics, Polish Academy of Sciences \\
ul. Radzikowskiego 152, PL-31342 Krak\'ow, Poland}}

\maketitle

\begin{abstract}
Popular parameterizations of the freeze-out conditions in relativistic heavy-ion
collisions are discussed. Similarities and differences between the blast-wave 
model and the single-freeze-out model, both used recently to 
interpret the RHIC data, are outlined. A non-boost-invariant extension of the
single-freeze-out model is proposed and applied to describe the recent BRAHMS data.
\end{abstract}


\thispagestyle{empty}

\section{The blast-wave model of Siemens and Rasmussen}
\label{sect:SR}

In 1979 Siemens and Rasmussen formulated a model describing the hadron production
in Ne + Na F reactions at the beam energy of 800 MeV per nucleon \cite{Siemens:1978pb}. 
The physical picture behind the model was that the fast hydrodynamic 
expansion of the produced hadronic matter leads to a sudden decoupling of hadrons and 
freezing of their momentum distributions, which retain their thermal character (although 
modified by the collective expansion effects) until the observation point. In their own 
words, Siemens and Rasmussen described the collision process as follows: "central 
collisions of heavy nuclei at kinetic energies of a few  hundred MeV per nucleon produce 
fireballs of hot, dense nuclear matter; such fireballs explode, producing blast waves 
of nucleons and pions". In this way, with Ref. \cite{Siemens:1978pb}, the concept of 
the blast waves of hadrons and the blast-wave model itself entered the field of
relativistic heavy-ion collisions.

Although the model of Siemens and Rasmussen was motivated by an earlier hydrodynamic
calculation by Bondorf, Garpman, and Zimanyi \cite{Bondorf:1978kz}, the results presented 
in Ref. \cite{Siemens:1978pb} were not obtained by solving the hydrodynamic equations but 
followed from the specific assumptions on the freeze-out conditions. The most important 
ingredient of the model was the spherically symmetric expansion of the shells of 
matter with constant radial velocity.  With an additional assumption about the times when 
such shells disintegrate into freely streaming hadrons (this point will be discussed in 
a greater detail in Sect. \ref{sect:rad}) Siemens and Rasmussen obtained the formula for 
the momentum distribution of the emitted hadrons~\cite{Siemens:1978pb}
\begin{equation} 
{dN \over d^3p} = Z \exp\left( -{\gamma E \over T} \right)
\left[ \left( 1 + {T \over \gamma  E} \right)
{\hbox{sinh} a \over a} - {T \over \gamma  E} \,\hbox{cosh}a \right].
\label{SR1}
\end{equation}
In Eq. (\ref{SR1}) $Z$ is a normalization factor, $E=\sqrt{m^2+p^2}$ denotes the
hadron energy,  $T$ is the temperature of the fireball (the same for all fluid shells), 
and $\gamma=(1-v^2)^{-1/2}$ is the Lorentz gamma factor with $v$ denoting the radial 
collective velocity (radial flow). A dimensionless parameter $a$ is defined by the 
equation 
\begin{equation}
a = {\gamma v p \over T}.
\label{a}
\end{equation}
Small values of $v$ (and $a$) correspond to small expansion rate and, 
as expected, a simple Boltzmann factor is obtained  from Eq. (\ref{SR1}) in the limit 
$v \to 0$,
\begin{equation} 
{dN \over d^3p} \to Z \exp\left( -{E \over T} \right).
\label{SR2}
\end{equation}
The fits to the data based on the formula (\ref{SR1}) gave $T$ = 44 MeV and 
$v$~=~0.373. Interestingly, the value of the radial flow $v$ turned out to be quite
large suggesting the strong collective behavior. This was an unexpected feature 
summarized by the authors with the statement: "Monte Carlo studies suggest that 
Ne + Na F system is too small for multiple collisions to be very important, thus, 
this evidence for a blast feature may be an indication that pion exchange 
is enhanced, and the effective nucleon mean free path shortened in 
dense nuclear matter". 

\section{Cooper-Frye formula}
\label{sect:CF}

Below we shall analyze the formal steps leading to Eq. (\ref{SR1}). Our starting point
is the expression defining the momentum distribution of particles as the integral 
of the phase-space distribution function $f(x,p)$ over the freeze-out hypersurface 
$\Sigma$, i.e., the renowned Cooper-Frye formula \cite{Cooper:1974mv},
\begin{equation}
E \, {dN \over d^3p} = {dN \over dy\, d^2p_\perp} =  
\int d^3\Sigma_\mu(x) p^\mu f(x,p).
\label{CF1}
\end{equation}
The three-dimensional element of the freeze-out hypersurface in Eq. (\ref{CF1}) may be 
obtained from the formula
\begin{equation}
d^3\Sigma_\mu = \varepsilon_{\mu \alpha \beta \gamma}
{d x^\alpha \over d\alpha }
{d x^\beta \over d\beta }
{d x^\gamma \over d\gamma } d\alpha d\beta d\gamma,
\label{d3sigma}
\end{equation}
where $\varepsilon_{\mu \alpha \beta \gamma}$ is the Levi-Civita tensor and
$\alpha,\beta,\gamma$ are the three independent coordinates introduced to 
parameterize the hypersurface.

We note that for systems in local thermodynamic equilibrium we have
\begin{equation}
E \, {dN \over d^3p} = \int d^3\Sigma_\mu(x) \, p^\mu 
f_{\rm eq} \left(u_\mu(x) \,p^\mu \right), 
\label{CF2}
\end{equation}
where the function $f_{\rm eq}$ is the equilibrium 
distribution function
\begin{equation}
f_{\rm eq}(E) = {1\over (2 \pi)^3}
\left[\exp\left({E - \mu \over T}\right) + \epsilon \right]^{-1}.
\label{eq}
\end{equation}
Here the case $\epsilon = +1\,(-1)$ corresponds to the Fermi-Dirac (Bose-Einstein)
statistics, and the limit $\epsilon \to 0$ yields the classical (Boltzmann) statistics. 
For a static fireball one finds
\begin{equation}
d^3\Sigma_\mu = (dV,0,0,0), \quad u_\mu = (1,0,0,0), 
\end{equation}
and Eq. (\ref{CF2}) is reduced to the formula
\begin{equation}
{dN \over d^3p} = V f_{\rm eq}(E),
\label{CF3}
\end{equation}
where $V$ is the volume of the system. Eq. (\ref{CF3}) agrees with Eq. (\ref{SR2}) 
in the classical limit if the normalization constant $Z$ is taken as
\begin{equation}
Z = {V \over (2\pi)^3} \exp\left({\mu \over T}\right).
\label{Z}
\end{equation}

\section{Spherically symmetric freeze-outs}
\label{sect:rad}

\begin{figure}[t]
\begin{center}
\subfigure{\includegraphics[angle=0,width=0.48\textwidth]{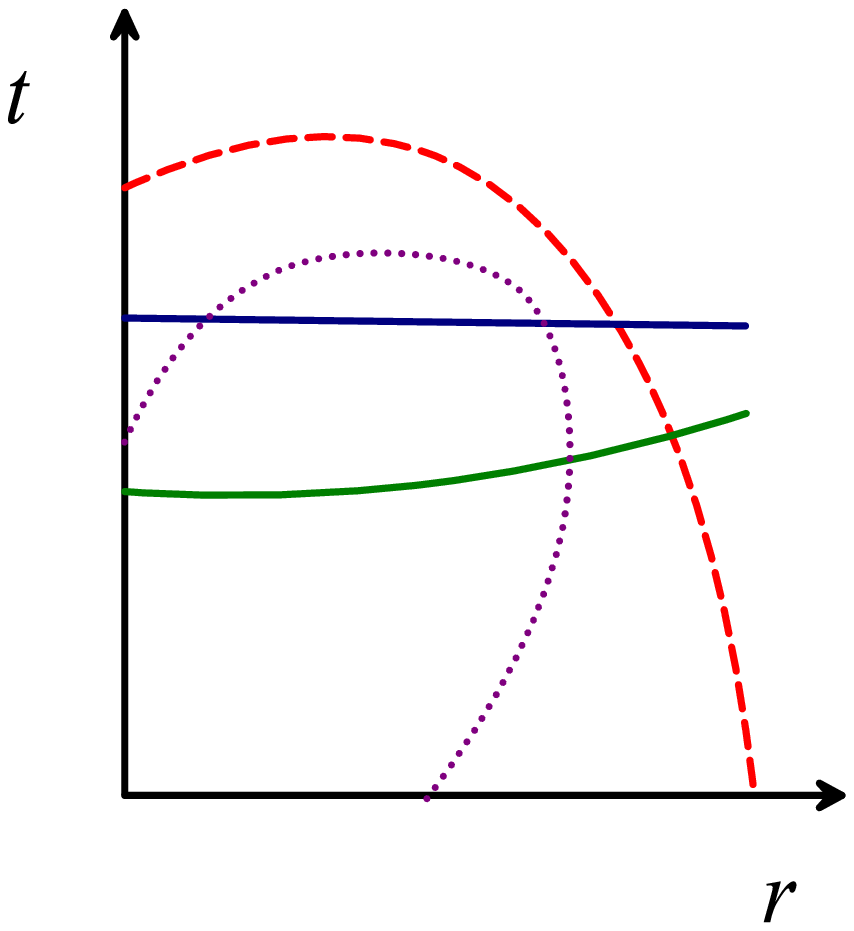}} 
\subfigure{\includegraphics[angle=0,width=0.48\textwidth]{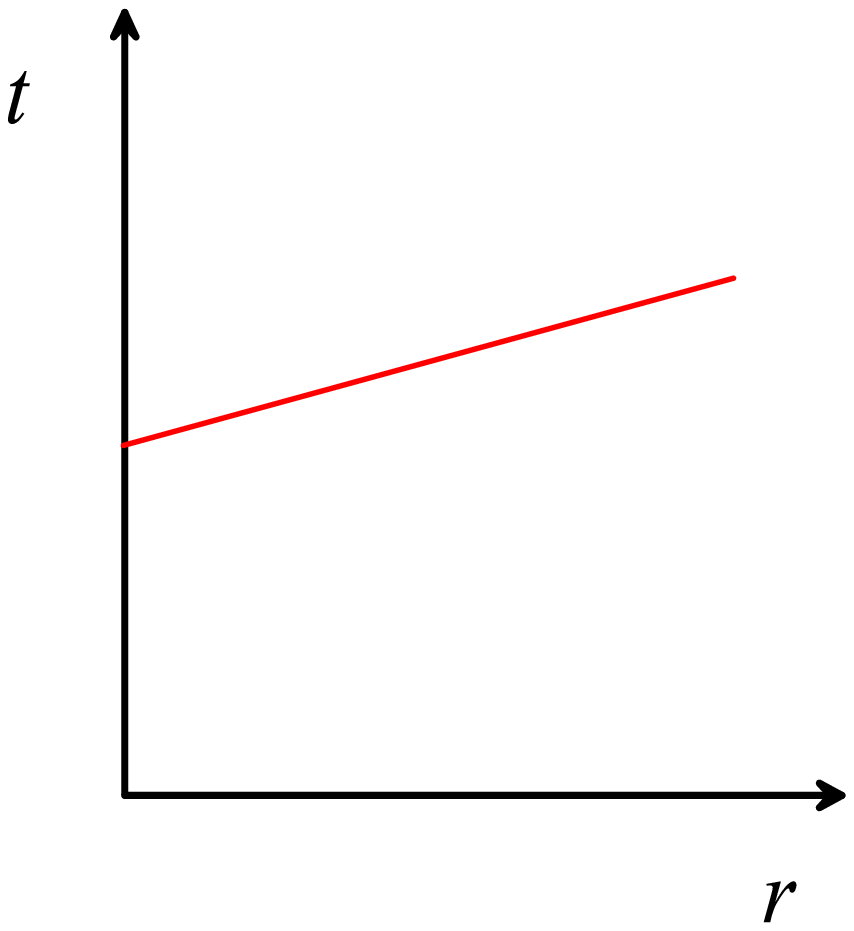}} \\
\end{center}
\caption{\small {Left: A priori possible, different freeze-out curves
in Minkowski space. The dotted and dashed lines describe the cases where both the
space-like and time-like parts are present. The solid lines describe the cases
where only the time-like part is present. Right: The (time-like) freeze-out curve 
assumed in the blast-wave model of Siemens and 
Rasmussen, compare Eq. (\ref{dtvdr}).}}
\label{fig:tr}
\end{figure}

For spherically symmetric freeze-outs it is convenient to introduce the
following parameterization of the space-time points on
the freeze-out hypersurface \cite{Rischke:1996em}
\begin{equation}
x^\mu = (t,x,y,z) = \left( t(\zeta), r(\zeta) \sin\theta \cos\phi, 
r(\zeta) \sin\theta \sin\phi, r(\zeta) \cos\theta \right).
\label{xmurad} 
\end{equation}
The freeze-out hypersurface is completely defined if a curve, i.e., the mapping $\zeta 
\longrightarrow \left(t(\zeta),r(\zeta)\right)$ in the $t-r$ space is given. This curve
defines the (freeze-out) times when the hadrons in the shells of radius $r$ stop 
to interact, see Fig. \ref{fig:tr}. The range of $\zeta$ may be always restricted to the 
interval: $0 \leq \zeta \leq 1$.  The three coordinates: $\phi \in [0,2 \pi], \theta \in
[0,\pi]$, and $\zeta \in [0,1]$ play the role of the variables $\alpha,\beta,\gamma$
appearing in Eq. (\ref{d3sigma}). Hence, the element of the spherically symmetric 
hypersurface has the form
\begin{eqnarray}
\! d^3\Sigma^\mu \!&=& \!
\left( r^\prime(\zeta), 
t^\prime(\zeta) \sin\theta \cos\phi,
t^\prime(\zeta) \sin\theta \sin\phi,
t^\prime(\zeta) \cos\theta \right)
r^2(\zeta) \sin\theta \,d\theta\,d\phi\,d\zeta, \nonumber \\
\label{d3sigmarad}
\end{eqnarray}
where the prime denotes the derivatives taken with respect to $\zeta$. Besides the
spherically symmetric hypersurface we introduce the spherically symmetric 
(hydrodynamic) flow
\begin{eqnarray}
u^\mu &=& \gamma(\zeta)\left( 1, 
v(\zeta) \sin\theta \cos\phi, 
v(\zeta) \sin\theta \sin\phi,
v(\zeta) \cos\theta  \right), 
\label{umurad1}
\end{eqnarray}
where $\gamma(\zeta)$ is the Lorentz factor, $\gamma(\zeta) = (1-v^2(\zeta))^{-1/2}$.
In a similar way the four-momentum of a hadron is parameterized as
\begin{equation}
p^\mu = \left[E, p \sin\theta_p \cos\phi_p, 
p \sin\theta_p \sin\phi_p,
p \cos\theta_p \right],
\label{pmurad}
\end{equation}
and we find the two useful expressions:
\begin{equation}
p \cdot u = \left(E - p v(\zeta) \cos\theta \right)\,\gamma(\zeta),
\label{purad}
\end{equation}
\begin{equation}
d^3\Sigma \cdot p = \left(E r^\prime(\zeta) 
- p t^\prime(\zeta) \cos\theta \right) r^2(\zeta) \sin\theta
\,d\theta \, d\phi \, d\zeta.
\label{Sigmaprad}
\end{equation}
We note that the spherical symmetry allows us to 
restrict our considerations to the special case $\theta_p = 0$.

In the case of the Boltzmann statistics, with the help  of Eqs. (\ref{CF1}), 
(\ref{purad}) and (\ref{Sigmaprad}), we obtain
the following form of the momentum distribution
\begin{equation}
E {dN \over d^3p} = \int\limits_0^1 { e^{-(E \gamma-\mu)/T} \over 2 \pi^2}
\left[ E \,{\hbox{sinh}a \over a} \, {dr\over d\zeta}
+ T \,{(\hbox{sinh}a - a \hbox{cosh}a) \over a \gamma v }
{dt \over d\zeta} \right] r^2(\zeta) d\zeta.
\label{dNd3prad1}
\end{equation}
Here $v, \gamma, r$ and $t$ are functions of $\zeta$, and the parameter $a$ is 
defined by Eq. (\ref{a}). The thermodynamic parameters $T$ and $\mu$ may also 
depend on $\zeta$. To proceed further we need to make certain assumptions
about the $\zeta$-dependence of these quantities. In particular, to obtain the model 
of Siemens and Rasmussen we assume that the thermodynamic parameters as
well as the transverse flow velocity are constant
\begin{equation}
T = \hbox{const}, \quad  \mu = \hbox{const}, \quad 
v = \hbox{const} \quad (\gamma = \hbox{const}  , \quad a = \hbox{const} ).
\end{equation}
Moreover, we should assume that
the freeze-out curve in the $t - r$ space satisfies the condition
\begin{equation}
dt = v \,dr, \quad t = t_0 + v r.
\label{dtvdr}
\end{equation}
In this case we obtain the formula 
\begin{equation}
{dN \over d^3p} = { e^{-(E \gamma-\mu)/T} \over 2 \pi^2}
\left[ \left(1+ {T \over \gamma E}\right) \,{\hbox{sinh}a \over a} \, 
- {T \over \gamma E} \, \hbox{cosh}a \,\right] 
\int\limits_0^1 r^2(\zeta) {dr \over d\zeta} d\zeta.
\label{dNd3prad2}
\end{equation}
Equation (\ref{dNd3prad2}) coincides with Eq. (\ref{SR1}) if we use Eq. (\ref{Z}) and
make the following identification
\begin{equation}
\int r^2(\zeta) {dr \over d\zeta} d\zeta = { r^3_{\rm max} \over 3}, \quad
V = {4 \over 3}  \pi r^3_{\rm max}.
\label{V}
\end{equation}
Note that the quantity $r_{\rm max}$ does not necessarily denote the maximum 
value of the radius of the system, see the dotted line on the left-hand-side of
Fig.~\ref{fig:tr}.

An interesting and perhaps unexpected feature of the model proposed
by Siemens and Rasmussen is the relation between the times and positions of
the freeze-out points, see Eq. (\ref{dtvdr}) illustrated on the right-hand-side 
part of Fig.~\ref{fig:tr}. Eq.~(\ref{dtvdr}) 
indicates that the fluid elements which are further away from the
center freeze-out later. Moreover, taking into account Eq.~(\ref{dtvdr}) 
in the formula (\ref{d3sigmarad}) we find that the four-vector describing the
hypersurface is parallel to the four-vector describing the flow, compare Eqs. 
(\ref{d3sigmarad}) and (\ref{umurad1}) giving $d^3\Sigma^\mu \sim u^\mu$
in this case. As we shall see the same features are assumed in the 
single-freeze-out model \cite{Broniowski:2001we,Broniowski:2001uk,
Broniowski:2002nf}.

It is worth to emphasize that in the hydrodynamic approach the $t-r$ freeze-out curves 
contain the space-like and time-like parts~\footnote{We use the convention that in the 
space-like (time-like) region the vector normal to the freeze-out curve is space-like 
(time-like).}. The treatment of the space-like parts leads to conceptual problems since 
particles emitted from such regions of the hypersurface enter again the system 
and the hydrodynamic description of such regions (combined with the
use of the Cooper-Frye formula) is inadeqate. Recently much work has been
done to develop a consistent description of the freeze-out process from the space-like 
parts \cite{Bugaev:2004kq,Csernai:2004pr}. 
However, very often only a quantitative argument is presented 
\cite{Teaney:2001av} that 
the contributions from the space-like parts are small and may be neglected
compared to the contributions from the time-like regions. The choice of
Siemens and Rasmussen seems to have anticipated such arguments.

\section{Boost-invariant blast-wave model of Schnedermann, Sollfrank, and Heinz}
\label{sect:binv}

The model presented above is appropriate for the low-energy scattering processes
where the two nuclei completely merge at the initial stage of the collision and further 
expansion of the system is, to large extent, isotropic. At higher energies such a picture 
is not valid anymore and, following the famous paper by Bjorken \cite{Bjorken:1983qr},  
the boost-invariant and cylindrically
symmetric models have been introduced to describe the collisons~\footnote{As is
discussed in greater detail below, the data delivered by the BRAHMS Collaboration
indicate that the systems produced at RHIC may be treated as boost-invariant only
in the limited rapidity range $-1 < y < 1$. Moreover, the assumption about the cylindrical
symmetry is valid only for the most central data.}.

The boost-invariance (symmetry with respect to the Lorentz transformations)
may be incorporated in the hydrodynamic equations, kinetic equations, and also in 
the modeling of the freeze-out process. In the latter case, the appropriate formalism
was developed by Schnedermann, Sollfrank, and Heinz \cite{Schnedermann:1993ws}.
The ansatz for the boost-invariant, cylindrically symmetric freeze-out hypersurface
has the form
\begin{equation}
x^\mu = (t,x,y,z) =
 \left( {\tilde \tau}(\zeta) \hbox{cosh } \alpha_\parallel,
\rho(\zeta) \cos\phi, \rho(\zeta) \sin\phi, 
{\tilde \tau}(\zeta) \hbox{sinh } \alpha_\parallel \right).
\label{xmubinv}
\end{equation}
Here, the parameter $\alpha_\parallel$ is the space-time rapidity. At $\alpha_\parallel=0$ 
the longitudinal coordinate $z$ is also zero and the variable ${\tilde \tau}(\zeta)$
coincides with the time coordinate $t$. Similarly to the spherical expansion discussed
in Sect. \ref{sect:rad}, the boost-invariant freeze-out hypersurface is completely 
defined if the freeze-out curve $\zeta \to \left( {\tilde \tau}(\zeta), \rho(\zeta) \right)$ 
is given. This curve defines the freeze-out times of the cylindrical shells with the radius 
$\rho$. Because of the boost-invariance it is enough to define this curve at $z=0$, since 
for finite values of $z$ the freeze-out points may be obtained by the Lorentz transformation.

The volume element of the freeze-out hypersurface is obtained from Eq.~(\ref{d3sigma}),  
\begin{equation}
d^3\Sigma^\mu = \left( {d\rho \over d\zeta} \hbox{cosh }\alpha_\parallel,
{d {\tilde \tau} \over d\zeta} \cos\phi,
{d {\tilde \tau} \over d\zeta} \sin\phi,
{d\rho \over d\zeta} \hbox{sinh }\alpha_\parallel \right)
\rho(\zeta){\tilde \tau}(\zeta) d\zeta d\alpha_\parallel d\phi.
\label{d3sigmabinv}
\end{equation}
Similarly to Eq. (\ref{xmubinv}) the boost-invariant four-velocity field has the 
structure
\begin{equation}
u^\mu = \hbox{cosh}\,\alpha_\perp(\zeta) \hbox{cosh}\,\alpha_\parallel
\left( 1 , \hbox{tanh}\,\alpha_\perp(\zeta) \cos\phi,
\hbox{tanh}\,\alpha_\perp(\zeta) \sin\phi,
\hbox{tanh}\,\alpha_\parallel \right). 
\label{umubinv}
\end{equation}
We note that  the longitudinal flow is
simply $v_z = \hbox{tanh}\, \alpha_\parallel = z/t$
(as in the one-dimensional Bjorken model), whereas the transverse
flow is $v_r = \hbox{tanh}\,\alpha_\perp(\zeta)$.

With the standard parameterization of the particle four-momentum 
in terms of rapidity $y$ and the transverse mass $m_\perp$,
\begin{equation}
p^\mu = \left(m_\perp \hbox{cosh} y, p_\perp \cos\varphi, p_\perp
\sin\varphi, m_\perp \hbox{sinh} y \right),
\label{pmubinv}
\end{equation}
we find
\begin{equation}
p \cdot u = m_\perp \hbox{cosh}(\alpha_\perp) 
\hbox{cosh}(\alpha_\parallel-y) -  p_\perp  
\hbox{sinh}(\alpha_\perp) \cos(\phi-\varphi),
\label{pubinv}
\end{equation}
and
\begin{equation}
d^3\Sigma  \cdot p = \left[m_\perp \hbox{cosh}(y-\alpha_\parallel) 
{d\rho \over d\zeta} - p_\perp \cos(\phi-\varphi) {d {\tilde \tau}
\over d\zeta} \right]
\rho(\zeta){\tilde \tau}(\zeta) d\zeta d\alpha_\parallel d\phi.
\label{sigmapbinv}
\end{equation}

For the Boltzmann statistics, with $\beta = 1/T$, the Cooper-Frye formalism gives
the following momentum distribution
\begin{eqnarray}
& & {dN \over dy d^2p_\perp}  = \nonumber \\
& & {e^{\beta \mu} \over (2\pi)^3}
\int\limits_0^{2\pi} d\phi 
\int\limits_{-\infty}^{\infty} d\alpha_\parallel \int\limits_0^1 d\zeta
\,\, \rho(\zeta) {\tilde \tau}(\zeta) \left[m_\perp 
\hbox{cosh}(\alpha_\parallel-y) {d\rho \over d\zeta} 
- p_\perp \cos(\phi-\varphi) {d{\tilde \tau} \over
d\zeta} \right] \nonumber \\
& & \times 
\exp\left[-\beta m_\perp \hbox{cosh}(\alpha_\perp) 
\hbox{cosh}(\alpha_\parallel-y)+ \beta p_\perp  
\hbox{sinh}(\alpha_\perp) \cos(\phi-\varphi) \right].
\label{bmdN1}
\end{eqnarray}
The form of Eq. (\ref{bmdN1}) shows explicitly that the distribution
$dN/(dy d^2p_\perp)$ is independent of $y$ and $\varphi$, in accordance with
our assumptions of the boost-invariance and cylindrical symmetry. 
The integrals over $\alpha_\parallel$ and $\phi$ in Eq. (\ref{bmdN1}) 
are analytic and lead to the Bessel functions $K$ and $I$,
\begin{eqnarray}
{dN \over dy d^2p_\perp} &=&
{e^{\beta \mu} \over  2 \pi^2} m_\perp
K_1 \left[\beta m_\perp \hbox{cosh}(\alpha_\perp) \right]  
I_0 \left[\beta p_\perp \hbox{sinh}(\alpha_\perp)\right]
\int\limits_0^1 d\zeta \,\, \rho(\zeta) {\tilde \tau}(\zeta) 
{d\rho \over d\zeta} 
\nonumber \\
&&  - {e^{\beta \mu} \over  2 \pi^2} p_\perp
K_0\left[\beta m_\perp \hbox{cosh}(\alpha_\perp) \right] 
I_1 \left[\beta p_\perp \hbox{sinh}(\alpha_\perp)\right]
\int\limits_0^1 d\zeta \,\, \rho(\zeta) {\tilde \tau}(\zeta) 
{d{\tilde \tau} \over d\zeta}. \nonumber \\
\label{KandI}
\end{eqnarray}
In the spirit of the blast-wave model of Siemens and Rasmussen we have assumed here
that the radial velocity is constant, $v_r = \hbox{tanh}\,\alpha_\perp(\zeta) 
= \hbox{const}$, otherwise the Bessel functions should be kept under the
integral over $\zeta$.

In order to achieve the simplest possible form of the model, the common practice 
is to neglect the second line of  Eq. (\ref{KandI}). 
This procedure means that one assumes implicitly the freeze-out condition 
$d{\tilde \tau} / d\zeta = 0 \, ({\tilde \tau} = \tau = \hbox{const})$. In this
case the boost-invariant blast-wave model is reduced to the formula
\begin{equation}
{dN \over dy d^2p_\perp} =
\hbox{const}\, m_\perp
K_1 \left[\beta m_\perp \hbox{cosh}(\alpha_\perp) \right]  
I_0 \left[\beta p_\perp \hbox{sinh}(\alpha_\perp)\right].
\label{K1I0a}
\end{equation}
where the constant has absorbed the factor $e^{\beta \mu} \tau \rho^2_{\rm max}/ (4 \pi^2)$.
Eq. (\ref{K1I0a}) forms the basis of numerous phenomenological analyses of the
transverse-momentum spectra measured at the SPS and RHIC 
\cite{Retiere:2003kf} energies.

\section{Resonances}

The main drawback of the formalism outlined above is that it neglects the effect
of the decays of hadronic resonances. Such an approach may be justified at lower energies
but should be improved at the relativistic energies where most of the light
particles are produced in the decays of heavier resonance states. The expressions giving
the rapidity and transverse momentum spectra of particles originating from two- and 
three-body decays of the resonances with a specified momentum distribution were worked 
out by Sollfrank, Koch, and Heinz \cite{Sollfrank:1990qz,Sollfrank:1991xm}. 
Their formulae may also be used to account for the feeding from the resonances 
in the blast-wave model, as already proposed in Ref.
\cite{Schnedermann:1993ws}. In other words, we wish to stress that the choice of the
freeze-out hypersurface and of the flow profile are elements {\it completely
independent}  of the treatment of the resonances. Both are important with the latter
being the basic ingredient in the calculation of particle abundances and the key
to success of thermal models.

At the SPS and RHIC energies it is important to include not only the decays of the
most common resonances such as $\eta, \rho, \omega, K^*$ or $\Delta$, but also of much
heavier states. Although their contributions are supressed by the Boltzmann factor,
their number increases strongly with the mass 
\cite{Broniowski:2000bj,Broniowski:2004yh}, 
hence their role can be easily underestimated. The effects of sequential decays of heavy 
resonances were first realized in statistical analyses of the
ratios of hadronic abundances/multiplicities 
(for recent results see \cite{Florkowski:2001fp,Braun-Munzinger:2001ip,Torrieri:2004zz,
Wheaton:2004qb}) 
which showed that the statistical models 
give a very good description of the data, provided most of the hadrons appearing in the
Particle Data Tables are included in the calculations. 

In order to discuss the role of the sequential decays of the resonances 
it is convenient to start with a general formalism giving the
Lorentz-invariant phase-space density of the measured particles \cite{Bolz:1992hc}
\begin{eqnarray}
&& n_{1 }\left( x_{1},p_{1}\right)  =  E_1 {dN_1 \over d^3p_1 d^4x_1 } =
\label{npix1p1} \nonumber \\ 
&& \int \frac{d^{3}p_{2}}{E_{p_{2}}}
B\left( p_{2},p_{1}\right) \int d\tau _{2}\Gamma _{2}e^{-\Gamma _{2}\tau
_{2}} \int d^{4}x_{2}\delta ^{\left( 4\right) }
\left( x_{2}+\frac{p_{2}\tau_{2}}{m_{2}}-x_{1}\right)...  \nonumber \\
&&\times \int \frac{d^{3}p_{N}}{E_{p_{N}}}B\left( p_{N},p_{N-1}\right) \int
d\tau _{N}\Gamma _{N}e^{-\Gamma _{N}\tau _{N}} \nonumber \\
&&   \hspace{5mm} \times \int d\Sigma _{\mu }\left(
x_{N}\right) \,p_{N}^{\mu }\,\,\delta ^{\left( 4\right) }\left( x_{N}+\frac{%
p_{N}\,\tau _{N}}{m_{N}}-x_{N-1}\right) f_{N}\left[ p_{N}\cdot u\left(x_{N}\right) 
\right]. \nonumber \\
\label{ornik1}
\end{eqnarray}
Here the indices $1,2, ..., N$ label hadrons in one chain of the sequential decays. 
The first resonance is produced on the freeze-out hypersurface and has the label $N$. 
The final hadron has the label $1$, for more details see 
\cite{Broniowski:2001uk,Broniowski:2002nf}.
The function $B(k,q)$ is the probability distribution for a
resonance with momentum $k$ to produce a particle with momentum $q$
in a two-body decay 
\begin{equation}
B(k,q) = {b \over 4 \pi p^*} 
\delta \left( {k \cdot q \over m_R}- E^* \right).
\label{Bkq}
\end{equation}
The function $B(k,q)$ satisfies the normalization condition
\begin{equation}
\int {d^3q \over E_q} B(k,q) = b,
\label{normB}
\end{equation}
where $b$ is the branching ratio for a given decay channel and $p^* (E^*)$
is the momentum (energy) of the emitted particle in the resonance's rest frame
(a generalization to three-body decays is straightforward and explained in
Refs. \cite{Florkowski:2001fp}).

Integration of Eq. (\ref{ornik1}) over all space-time positions gives
the formula for the momentum distribution
\begin{eqnarray}
&& E_{p_1} {dN_1 \over d^3 p_1} = \int d^{4}x_{1}\,n_{1}
\left(x_{1},p_{1}\right) = \label{npip1} \nonumber \\
&&\int \frac{d^{3}p_{2}}{E_{p_{2}}}B\left( p_{2},p_{1}\right)...\int 
\frac{d^{3}p_{N}}{E_{p_{N}}}B\left( p_{N},p_{N-1}\right) 
\int d\Sigma_{\mu }\left( x_{N}\right) \,p_{N}^{\mu }\,\,f_{N}
\left[ p_{N}\cdot u\left(x_{N}\right) \right]. \nonumber \\
\label{ornik2}
\end{eqnarray}

Equation (\ref{ornik2}) serves as the starting point to prove that
for constant values of the thermodynamic parameters on the 
freeze-out hypersurface the ratios in the full phase-space
($4 \pi$) are the same as in the local fluid elements. In this way, 
a connection between the measured ratios and the local thermodynamic parameters
is obtained \cite{Heinz:1998st}. One may also check that for the boost-invariant systems
it is enough to consider the ratios at any value of the rapidity to infer the 
values of the thermodynamic parameters \cite{Broniowski:2001we,Broniowski:2002nf}.

The experimental RHIC data show, however,  that the rapidity distributions 
are of Gaussian shape \footnote{This feature has revived the interest in the 
Landau hydrodynamic model \cite{Landau:1953gs,Belenkij:1956cd}, see, for 
example, Ref. \cite{Steinberg:2004vy}.}
and the thermodynamic parameters 
vary with rapidity (the measured ${\bar p}/p$ ratio depends on $y$), hence, the
system created at RHIC is, strictly speaking, not boost-invariant. In this situation 
the relation between the measured ratios and thermodynamic parameters is not obvious. 
Fortunately, the RHIC data show also a rather flat rapidity distribution and
constant ratios in the rapidity range $-1 < y < 1$
\cite{Bearden:2003fw,Bearden:2004yx}. In this region (the central part of the
broad Gaussian) the system to a good approximation
may be treated as boost-invariant and the standard analysis
of the ratios may be performed to obtain the thermodynamic parameters at $y=0$.

\section{Single-freeze-out model}

The analysis of the ratios of hadron multiplicities measured at RHIC
gives a typical temperature of 170 MeV. On the other hand, the analysis
of the spectra based on Eq. (\ref{K1I0a}) gives a lower temperature of about 
100 - 140 MeV. Such a situation was observed already at the SPS energies, 
which motivated the introduction of the concept of two different freeze-outs. 

Certainly, if the spectra contain important contributions from high lying
states, the value of $T$ obtained from the blast-wave formula fitted to
the spectra cannot be interpreted as the temperature of the system 
in the precise thermodynamic sense. First, the contributions from the
resonances (feeding mostly the low-momentum region) should be subtracted
from the spectra of light hadrons, giving the insight to the properties
of the primordial particles. Using other words, we may argue that the
calculation of the ratios should include the same number of the resonances
as the corresponding calculation of the spectra.

An example of such a calculation is the single-freeze-out
model formulated in Refs. \cite{Broniowski:2001we,Broniowski:2001uk}. In this model
the decays of the resonances as well as the transverse flow change the spectra
of the primordial particles in such a way that it is possible to describe well
the spectra and the ratios with a single value of the temperature. The basic effect
here is that the hadronic decays lead to effective cooling of the spectra.

Similarly to the original blast-wave models discussed above, the single freeze-out model
assumes a certain form of the freeze-out hypersurface in the Minkowski space. 
In this case it is defined by the constant value of the proper time
\begin{equation}
\tau = \sqrt{t^2-r^2_x-r^2_y-r^2_z} = {\rm const}.
\label{tau}
\end{equation}
The transverse size of the system is defined by the parameter $\rho_{\rm max}$,
\begin{equation}
\rho=\sqrt{r_x^2+r_y^2}, \quad
\rho < \rho_{\rm max},
\label{rhomax}
\end{equation}
and the velocity field at freeze-out is taken in the Hubble-like form
\footnote{For a recent attempt to connect the parameterization (\ref{tau})
- (\ref{hubflow}) with hydrodynamic calculation see Ref. \cite{Chojnacki:2004ec}.}
\begin{equation}
u^\mu = \frac{x^\mu}{\tau} = \frac{t}{\tau} \left(1, 
\frac{x}{t}, \frac{y}{t}, \frac{z}{t}\right).
\label{hubflow}
\end{equation}
The natural parameterization of the freeze-out hypersurface has the form
\begin{eqnarray}
t &=&\tau \cosh \alpha _{\parallel }\cosh \alpha _{\perp },\quad z=\tau
\sinh \alpha _{\parallel }\cosh \alpha _{\perp },  \nonumber \\
x &=&\tau \sinh \alpha _{\perp }\cos \phi ,\quad y =\tau \sinh \alpha
_{\perp }\sin \phi, 
\label{txyz}
\end{eqnarray}
which may be considered as the special case of the formula (\ref{xmubinv}).
Eq. (\ref{d3sigma}) leads to the following
expression defining the volume element
\begin{equation}
d\Sigma^\mu(x) = u^\mu(x)\, \tau ^{3} \, {\rm sinh}(\alpha _{\perp})
{\rm cosh}(\alpha _{\perp}) \, d\alpha _{\perp}
d\alpha _{\parallel } d\phi.
\label{d3sigmaumu}
\end{equation}

A very important feature of the choice (\ref{tau}) - (\ref{hubflow}) is that 
the volume element is proportional to the four-velocity field. This feature 
holds also in the
model of Siemens and Rasmussen. In this case the treatment of the resonance
is very much facilitated. In particular, Eq. (\ref{ornik2}) may 
be rewriten in the form
\begin{eqnarray}
E_{p_1} {dN_1 \over d^3 p_1}\!\!\!\!&=&\!\!\!\!\int d\Sigma \left(
x_{N}\right) \int \frac{d^{3}p_{2}}{E_{p_{2}}}
B\left( p_{2},p_{1}\right) \nonumber \\
& & ...
\int \frac{d^{3}p_{N}}{E_{p_{N}}}B\left( p_{N},p_{N-1}\right) p_{N}
\,\cdot u\left( x_{N}\right) \,f_{N}\left[ p_{N}\cdot u\left( x_{N}\right)
\right]  \nonumber \\
&=&\int d\Sigma \left( x_{N}\right)
p_{1}\,\cdot u\left( x_{N}\right)
\,f_{1}\left[ p_{1}\cdot u\left( x_{N}\right) \right], 
\label{ornik3}
\end{eqnarray}
where we have introduced the notation
\begin{eqnarray}
& & p_{i-1}\,\cdot u\left( x_{N}\right) \,f_{i-1}\left[ p_{i-1}\cdot
u\left( x_{N}\right) \right] \label{trsim} 
\nonumber \\& & =\int \frac{d^{3}p_{i}}{E_{p_{i}}}B\left(
p_{i},p_{i-1}\right) p_{i}\,\cdot u\left( x_{N}\right) \,f_{i}\left[
p_{i}\cdot u\left( x_{N}\right) \right]. 
\label{acta1}
\end{eqnarray}
In the local rest frame, the iterative procedure defined by Eq. (\ref{acta1}) 
becomes a simple one-dimensional integral transform 
\begin{equation}
f_{i-1}\left( q\right) = 
\frac{b m_R}{2 E_q p^\ast q} \int_{k_{-}(q)}^{k_{+}(q)} dk\, k \, 
f_i\left( k\right),
\label{acta2}
\end{equation}
where $k_\pm (q) = m_R | E^* q \pm p^* E_q |/m_1^2$. Eqs. (\ref{acta1}) and
(\ref{acta2}) allow us to deal with a very large number of decays in the very efficient
way, very similar to that used in the calculation of the hadron abundances.

\begin{figure}[b]
\begin{center}
\subfigure{\includegraphics[angle=0,width=0.95\textwidth]{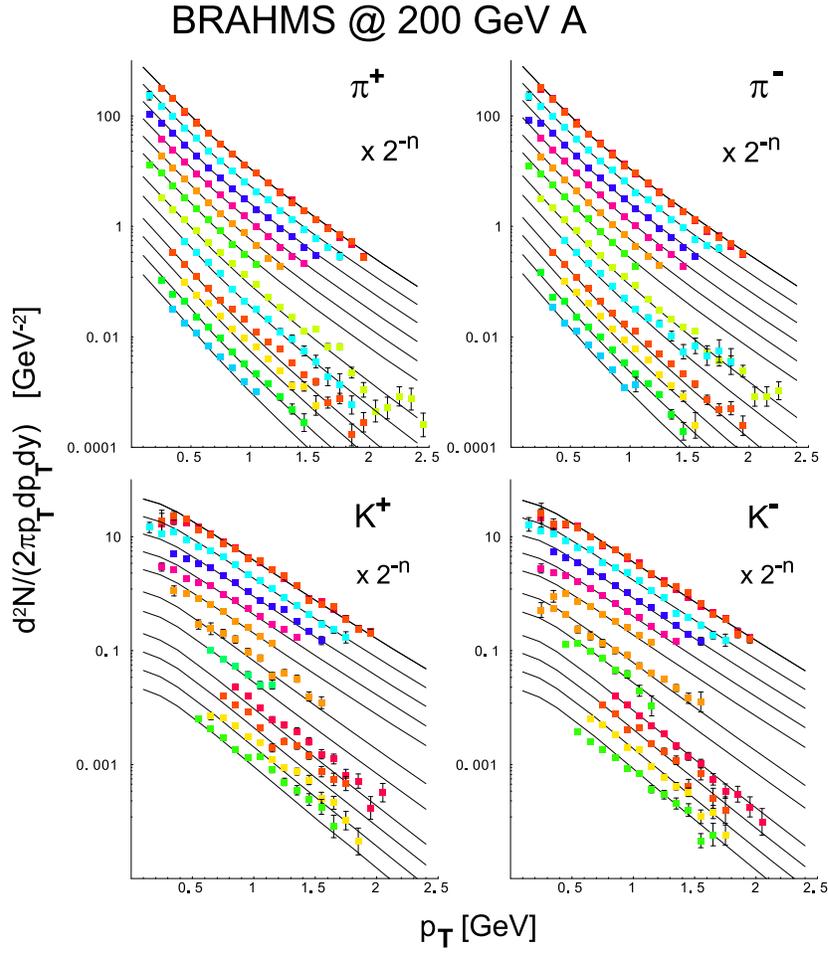}} 
\end{center}
\caption{ The single-freeze-out model fit to the transverse-momentum
spectra measured by the BRAHMS Collaboration for different values
of the rapidity \cite{Bearden:2004yx}. The 
successive curves correspond, from top to bottom, to
the rapidity bins centered at: 
$y=-0.05, \,0.05, \,0.5, \,0.7, \,0.9, \,1.1, \,1.3, 
\,2.2, \,2.5, \,3.05, \,3.15, \,3.25, \,3.35, \,3.53$ \cite{Bearden:2004yx}. 
The data in the third, fourth etc. bin are subsequently divided by factors
of 2, which is indicated by the $2^{-n}$ label.}
\label{fig:brpt}
\end{figure}

\section{Non boost-invariant single-freeze-out model}

The model described above may be generalized to the non boost-invariant
version in the minimal way by the modification of the system boundaries.
Introducing a dependence of the transverse size 
on the longitudinal coordinate $z$ (or $\alpha_\parallel$), we break explicitly
the assumption of the boost-invariance. At the same time, however, the
local properties of the hypersurface and flow remain unchanged 
allowing us to treat the resonances in the same simple way as described in
the previous Section.

Since the measured rapidity distributions are approximately gaussian, it is natural to
start with the gaussian ansatz for the dependence of the transverse size on
the parameter $\alpha_\parallel$ and restrict the region of the integration over
$\alpha_\perp$ to the interval
\begin{equation}
 0 \leq \alpha_\perp \leq 
\alpha_\perp^{\rm max} \exp[- \alpha^2_\parallel/(2 \Delta^2)].
\end{equation}
The original boost-invariant version is recovered in the limit 
$\Delta \longrightarrow \infty$. 
Using the values of the thermodynamic parameters obtained from the
boost-invariant version of the model applied to the hadronic ratios measured
at midrapidity ($T$ = 165.6 MeV and $\mu_B = 28.5$ MeV), we are left with
three extra parameters ($\tau$, $\rho_{\rm max} = \tau \, \hbox{sinh} \,
\alpha_\perp^{\rm max}$,
and $\Delta$), which should be fitted to the $p_\perp$-spectra collected
at different values of the rapidity.

\begin{figure}[t]
\begin{center}
\subfigure{\includegraphics[angle=0,width=0.7\textwidth]{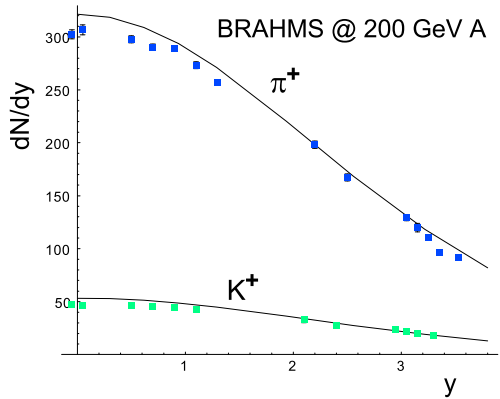}} 
\end{center}
\caption{Comparison of the measured rapidity distributions \cite{Bearden:2004yx}
with the results of the single-freeze-out model. The theoretical
curves were obtained by the integration of the spectra shown in Fig. \ref{fig:brpt}.  }
\label{fig:bry}
\end{figure}

The result of such a fit to the available BRAHMS data on $\pi^+$, $\pi^-$, $K^+$, and 
$K^-$ production  are shown in Fig. \ref{fig:brpt}.  The optimal values of the
parameters found in the fit are: $\tau = 8.33$ fm, $\alpha_\perp^{\rm max} = 0.825$, 
and $\Delta = 3.33$. One can see that the model reproduces the data very well in a
wide range of the transverse-momentum and rapidity.
In Fig. \ref{fig:bry} we show the model rapidity distributions compared to the data.
Small discrepancies (of about 10\%) between the model and the data may be seen
for the pions at $y=0$. Note that the comparison of the rapidity distributions in Fig. 
\ref{fig:bry} is done with a linear scale; definitely, small discrepancies 
may be expected for a such simplified description of the freeze-out.

It should be emphasized that the non-boost invariant version of the model presented 
above is not capable of describing correctly the ${\bar p}/p$ ratio. 
In the present framework this requires
an introduction of the rapidity dependence of the baryon chemical potential.

\section{Conclusions}

We have discussed several parameterizations of the freeze-out conditions in the 
relativistic heavy-ion collisions. We have argued that the single freeze-out model
used to describe the RHIC data is a natural development of the blast-wave models
worked out, among others,  by Siemens, Rasmussen, Heinz, Schnedermann, and Sollfrank.
The main advantage of the single-freeze-out model is that it includes all well established
resonance decays, allowing us to treat the chemical and thermal freeze-out 
as essentially one phenomenon. In this respect, the single-freeze-out model is very
similar to the original blast-wave model. Further similarities concern the shape
of the freeze-out hypersurface (only the time-like parts are considered) and
the strict use of the Cooper-Frye formula. Due to the limited space, we have not
discussed here the variety of models where, instead of the Cooper-Fry formula, the
so-called emission functions are introduced and modeled. An example of such an a
approach is the Buda-Lund model \cite{Csanad:2004mm}. 

One of us (WF) acknowledges clarifying discussions with Jan Rafelski and
Giorgio Torrieri.

\end{document}